# Roughness-Dependent Tribology Effects on Discontinuous Shear Thickening

Chiao-Peng Hsu,[1,2] Shivaprakash N. Ramakrishna,[2]
Michele Zanini,[1] Nicholas D. Spencer,[2] and Lucio Isa[1, *]

[1]*Laboratory for Interfaces, Soft Matter and Assembly,*
*Department of Materials, ETH Zurich, Zurich, Switzerland.*
[2]*Laboratory for Surface Science and Technology,*
*Department of Materials, ETH Zurich, Zurich, Switzerland.*

* lucio.isa@mat.ethz.ch

Surface roughness affects many properties of colloids, from depletion [1] and capillary interactions [2], to their dispersibility [3] and use as emulsion stabilizers [4]. It also impacts particle-particle frictional contacts, which have recently emerged as being responsible for the discontinuous shear thickening (DST) of dense suspensions [5–17]. Tribological properties of these contacts have been rarely experimentally accessed [6, 15, 17, 18], especially for non-spherical particles. Here, we systematically tackle the effect of nanoscale surface roughness by producing a library of all-silica, raspberry-like colloids [19] and linking their rheology to their tribology. Rougher surfaces lead to a significant anticipation of DST onset, both in terms of shear rate and solid loading. Strikingly, they also eliminate continuous thickening. DST is here due to the interlocking of asperities, which we have identified as "stick-slip" frictional contacts by measuring the sliding of the same particles via lateral force microscopy (LFM). Direct measurements of particle-particle friction therefore highlight the value of an engineering-tribology approach to tuning the thickening of suspensions.

Shear thickening (ST) is an intriguing rheological phenomenon, by which the viscosity $\eta$ of a concentrated particulate suspension increases upon increasing shear rate $\dot{\gamma}$ (or shear stress $\sigma$) above a critical value [9, 20]. Viscosity can either gradually increase (continuous shear thickening - CST) or diverge at a critical shear rate (discontinuous shear thickening - DST). In the most extreme cases, the material can even fully solidify under flow (shear jamming) [21]. DST can either be desirable, e.g. in impact-absorption applications [22], or highly detrimental, e.g. leading to clogging and pumping failures in the processing of dense slurries.

Although well characterized at the macroscale, the microscopic mechanisms governing the origins of ST are still not fully understood [23]. Hydrodynamic interactions play an essential role in the viscosity increase in CST [24–28], but alone they cannot predict the viscosity divergence in DST associated with a positive first normal stress difference $N_1$ [8, 16, 29]. In contrast, dilatancy ($N_1 > 0$) is a well-known feature of dense, frictional granular materials, reflecting the formation of anisotropic force-chain networks under shear [30–32]. This analogy has generated a growing consensus between theory [7], simulations [5, 8, 12–15] and experiments [6, 9–11, 16, 17], which have connected DST to the formation of stress-bearing structures of particles making solid-solid frictional contacts when hydrodynamic lubrication films break at high shear.

In spite of this significant body of work, often the input friction coefficients in numerical simulations do not reflect realistic values and only very few studies have actually attempted to measure the frictional properties of particles experimentally, either macroscopically [15] or microscopically [6, 17], and have been limited to smooth spheres [18]. Shear-thickening systems in applications, such as cementitious slurries, or the paradigmatic case of cornstarch suspensions, often comprise irregularly shaped particles. The geometry of contact is an essential component to describe frictional interactions, but, to date, only few studies have investigated the effect of particle topography, i.e. surface roughness, on ST. In general, higher roughness was shown to lead to the reduction of the onset rate and stress for DST and a sign change in $N_1$, from negative to positive, but no connection was made to the microscopic tribological properties of the particles [16, 29].

In this work, by experimentally studying the nanotribology of model silica colloids with tunable roughness, we demonstrate the existence of a direct link between particle topography, nanoscale friction and macroscopic DST. Engineering of the surface design of the particles allows us to control both the critical rate and the critical solid loading for DST, driven by an interlocking mechanism that is qualitatively different from the case of smooth particles.

We fabricate our model rough colloids by electrostatic adsorption of silica nanoparticles ("berries") onto larger silica colloids ("cores"). We then grow a controlled smoothing layer via a sol-gel route, creating all-silica raspberry-like particles, as shown in Figure 1 (a) [4]. Surface roughness can be tuned by independently choosing the size of the berries (12-39 nm) as well as by adjusting the thickness of the smoothing layer (10-15 nm) (see Figure 1 (b)-(g)). Surface roughness is then characterized by atomic force microscopy (AFM), and we extract a dimensionless roughness parameter $h/d$, calculated as the ratio between the average asperity height and the average inter-asperity separation, as shown in Figure 1 (h). We synthesize a library of raspberry-like particles with $\approx 0 < h/d < 0.53$, covering a broad roughness range from the smooth cores to the roughest raspberry. (See Methods for further details.)

We first quantify the role of surface roughness on the maximum packing fraction $\phi_m$ of the particles in a sedimentation/compressive rheology test. This quantity represents the limit at which the suspension can be processed, i.e. the volume fraction for which the suspensions jams at vanishingly small rates. Our previous work showed that $\phi_m$ is directly correlated to the inter-particle friction coefficient [6]. As opposed to the case of non-Brownian particles, $\phi_m$ can slowly evolve with time due to the combined effect of thermal fluctuations and sedimentation (more details in Methods and Figure S1). $\phi_m$ can be estimated by measuring the height of the sediment starting from a dilute suspension of known solid loading (Figure 2 (a)). Figure 2 (b) shows that the sediment height increases with the initial volume fraction, as expected. Rougher colloids present $\phi_m$ values that are clearly lower than those of smooth colloids (Figure 2 (c)). This indicates that rougher particles, i.e. with higher $h/d$ values, jam earlier during sedimentation and, as a result, the sediment is looser. Remarkably, particles with $h/d = 0.53$ jam under centrifugation for solid loadings as low as 44.5%, indicating that roughness has has a dramatic impact on DST and can be very effectively used to engineer the suspension's rheological response.

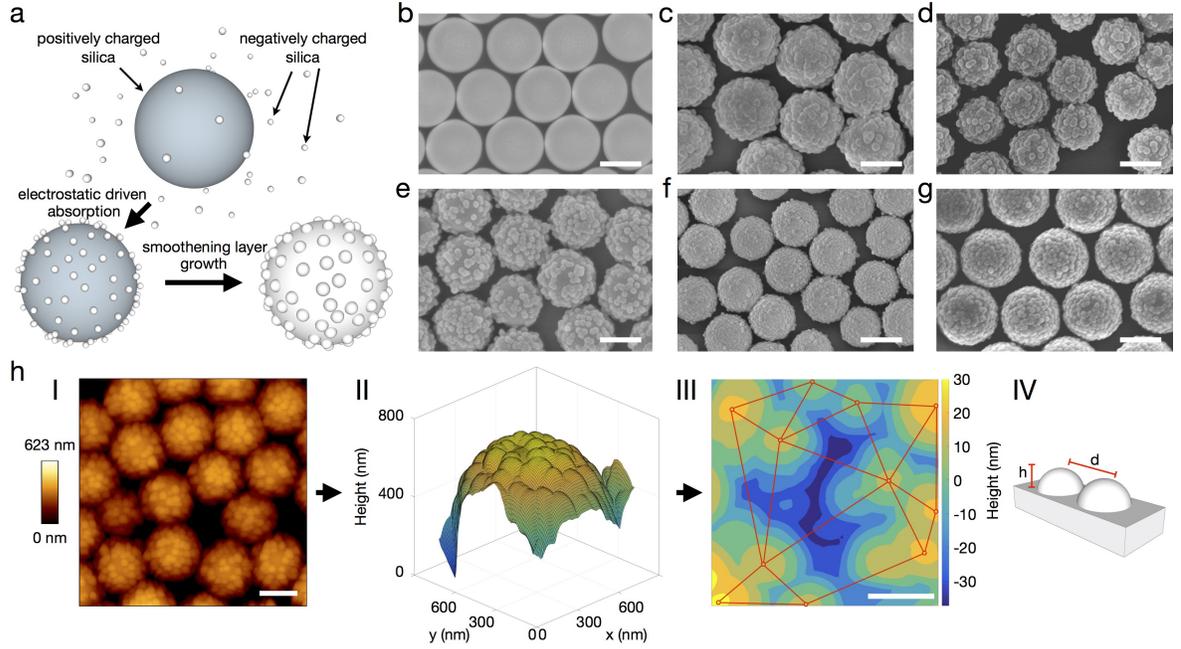

Figure 1. Fabrication and characterization of smooth (SM) and rough (RB) particles. (a) Schematics of the fabrication of raspberry-like particles. SEM images of (b) SM_0, (c) RB_0.25, (d) RB_0.31, (e) RB_0.36, (f) RB_0.45, (g) RB_0.53, scale bars = 500 nm. The numbers represent the value of $h/d$ for each batch. (h) AFM image of a rough particle monolayer (I), scale bar = 500 nm, surface topography image of a single rough particle (II), and contour plot of the central region of the flattened surface of the same particle (III), scale bar = 100 nm. The red circles identify the center of asperities and the red lines show the distance between asperities. Schematic definition of the roughness parameter $h/d$ (IV).

In fact, smooth colloids (SM_0, Figure 2 (d)) start to display CST for $\phi > 51$ % and only exhibit DST behavior at $\phi = 58$ %, which is very close to their measured $\phi_m$ of 59.2 %. The first normal stress difference $N_1$ remains negative between 48 % to 58 % during CST, while it switches sign in at the onset of DST which is characteristic of frictional dilatant flows. Rough colloids, on the other hand, show a qualitatively different behavior. Raspberry-like particles with $h/d = 0.53$ do not show any appreciable CST, but immediately discontinuously thicken, even for values of $\phi$ significantly lower than their $\phi_m$ (Figure 2 (e)), and the onset of DST shifts to lower $\dot{\gamma}$ with increasing $\phi$. It is also worth noting that the critical rate varies over almost two decades, compared to a much narrower window for the smooth colloids. Correspondingly, the viscosity increase is always associated with a positive $N_1$, indicating that DST is driven by frictional forces. The visual proof of the different ST response for smooth and rough particle suspensions is demonstrated in Video S1-8, showing ball-impact tests. Finally, Figure 2 (f) shows that, at the same solid loading of $\phi = 48$ %, rough colloids with different roughnesses exhibit DST, while smooth colloids do not thicken at all. The critical DST shear rate depends on the distance from $\phi_m$: The closer $\phi$ is to $\phi_m$, the lower the observed critical shear rate.

In order to account for these rheological observations, we turn to studying the microscopic particle-to-particle friction. These measurements are carried out by means of lateral force microscopy (LFM), where smooth and rough colloids are attached onto tip-less cantilevers (Figure S3) and scanned over planar substrates of varying roughness (roughness gradients), as shown in Figure 3 (a) and (b) (see Methods for further details). The substrates are produced by a process analogous to the synthesis of the rough colloids, to provide representative realistic counter-surfaces (see Methods and Figure S2). The LFM results from sliding a RB_0.53 probe over a roughness gradient with 22 nm-high asperities are shown in Figure 3 (c). (See Figure S4-7 for the friction results of all other particles.) Starting from the smooth end of the sample (rightmost curve - magenta), we observe a very narrow friction loop, i.e. a small difference in the lateral force signals between trace and retrace of the same scan on the substrate, indicative of a low friction coefficient. As soon as the area density of asperities increases, distinctive spikes arise in the friction-loop scans (cyan curve). These are typical of stick-slip frictional behavior; during scanning, when the probe meets an asperity, the lateral force increases steeply as the probe is locally stuck and then rapidly slides as the asperity is overcome. The frequency of the stick-slip events increases with increasing roughness (Figure 3 (c) from right to left), which corresponds to higher dissipation during scanning, and hence to an increase in the friction coefficient $\mu$ (Figure 3 (d)). The nature of the frictional interactions between rough surfaces also motivates our choice to describe surface roughness by the parameter $h/d$, since the stick-slip events are determined by the asperities' amplitude and periodicity [33], which are also the parameters we tune in the fabrication of our colloids.

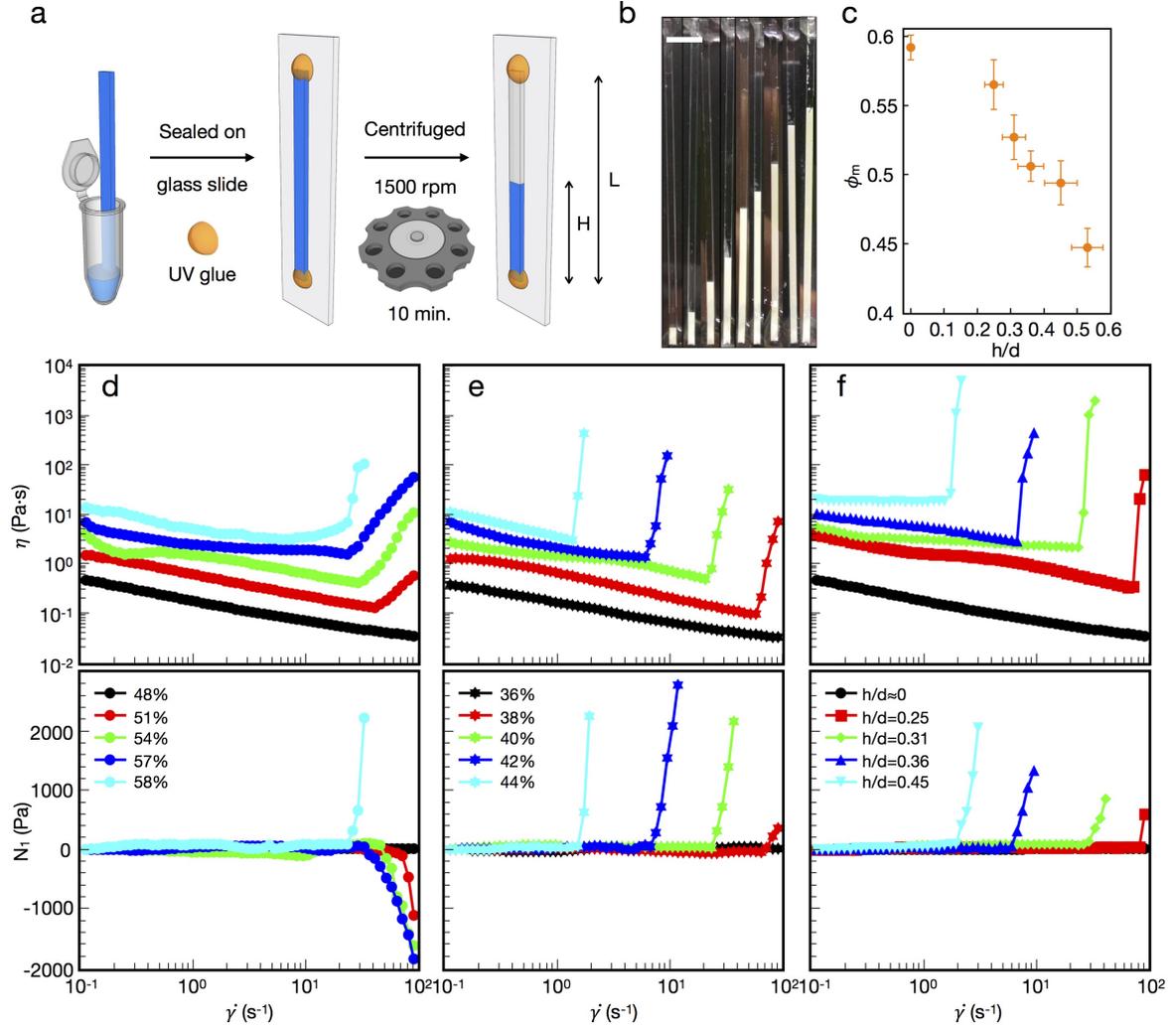

Figure 2. Results of compressive and shear rheology experiments. (a) Schematics of the centrifugation experiments, H is the height of the sediment and L is the length of the capillary. (b) Images of particle suspensions (SM_0) after centrifugation, the initial volume fraction $\phi_i$ increases from 5.7% (left) to 51.6% (right), scale bar = 5 mm. (c) $\phi_m$ of the colloidal suspensions with different surface roughness expressed in terms of $h/d$. Flow curves (top) and $N_1$ (bottom) of (d) smooth colloids SM_0 ● at different $\phi$ (48 % to 58 %) (e) rough colloids RB_0.53 ★ at different $\phi$ (36 % to 44 %) (f) smooth colloids SM_0 ●, rough colloids RB_0.25 ▼, rough colloids RB_0.31 ▲, rough colloids RB_0.36 ◆, rough colloids RB_0.45 ■ at $\phi = 48$ %

Interestingly, smooth and rough probes sliding on surfaces with increasing $h/d$ roughness give rise to different frictional dissipations (Figure 3 (e)). Generally, $\mu$ increases with surface roughness, but in a low roughness regime ($h/d < 0.3$), there are fewer asperities on the substrate and $\mu$ is mainly determined by the contact area of the two sliding surfaces rather than by stick-slip events. Rough probes contact the substrate via the asperities on their surfaces, resulting in smaller contact area and hence lower $\mu$ compared to smooth probes. Conversely, in a high roughness regime ($h/d > 0.3$), the density of asperities on the surface increases, so that stick-slip events are the main contribution to friction forces. The asperities on raspberry-like particles interlock with the asperities on the substrates, leading to higher $\mu$ values than those measured for smooth particles. Figure 3 (e) ultimately shows that there is a direct correlation between surface roughness and friction coefficient, which univocally depends on $h/d$ of the two surfaces.

The univocal dependence of both $\mu$ and $\phi_m$ on $h/d$ makes it possible to obtain a direct relation between the first two quantities, linking microscopic tribological properties with macroscopic rheological ones, as shown in Figure 3 (f). By plotting the friction coefficients of our smooth and rough particles against surfaces with the same $h/d$ versus a normalized maximum packing fraction $(\phi^{RCP} - \phi_m)/\phi^{RCP}$, where $\phi^{RCP} = 0.64$ is the random-close packing of monodisperse frictionless spheres, we see that the higher the interparticle friction coefficient, the lower the maximum packing fraction at which the material can be processed before DST occurs at vanishingly small rates. Moreover, the nanotribological measurements have also shed light on the nature of the qualitative difference in the ST behavior between smooth and rough particles. For the latter, as soon as a hydrodynamic lubrication films break, asperities interlock, giving rise to the formation of force chains and dilatant ($N_1 > 0$)

DST, while smooth particles experience standard sliding friction.

Finally, this correlation allows us to engineer the macroscopic rheological response, i.e. the $\phi_m$ of the suspension, by changing its nanoscopic tribological properties, i.e. the $\mu$ between the colloidal particles. To examine this concept, we perform the sedimentation experiments on mixed colloid suspensions obtained by introducing increasing fractions of smooth particles into suspensions of rough colloids (Figure 4). Remarkably, by adding an amount of smooth particles as low as 3.3 vol%, $\phi_m$ increases by more than 6 % and the onset rate for DST at $\phi = 0.44$ increases by almost two decades. Increasing the percentage of smooth colloids further, $\phi_m$ of the mixture tends towards the $\phi_m$ of the suspension of smooth colloids, but the biggest effect is seen within the first 10%. This strong effect is due the fact that smooth particles act like lubricants in the suspension by preventing strong interlocking between rough particles. The formation of stress-bearing network is delayed or reduced, leading to denser packings before DST (note that the mixtures still exhibit DST and positive $N_1$ ).

In conclusion, our results clearly confirm that there exists a strong link between the tribology of inter-particle contacts and the rheology of DST suspensions. The frictional properties greatly depend on the contact geometry, and surface roughness has emerged as an essential design parameter for the thickening response. We have, for instance, shown that one can increase the solid loading and delay undesired shear thickening by introducing a small amount of more lubricious particles into the system, which could be of interest for slurry processing, for example. Conversely, increasing surface roughness enables a great reduction of the volume fraction, while retaining very strong thickening but having lower viscosities in the unthickened region of the flow curve, which could be of interest in fluid materials for vibration or impact absorption. As the importance of tribology in thickening fluids is increasingly becoming more widely accepted, we expect many exciting opportunities for nanoscale surface design.

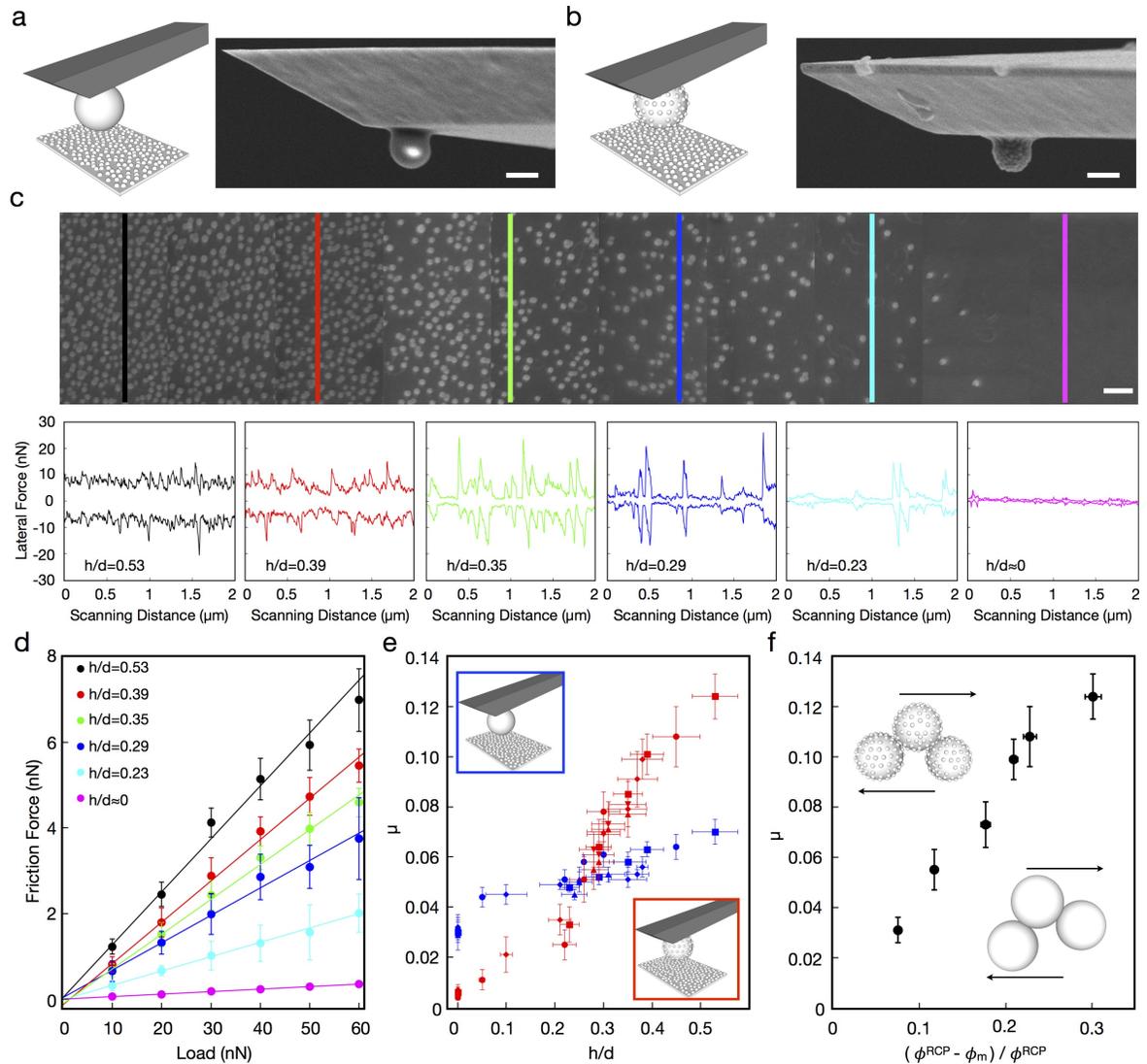

Figure 3. Friction measurements on model rough substrates. (a) Schematics of a smooth probe on a rough sample, and SEM image of a smooth colloidal probe, scale bar = 500 nm. (b) Schematics of a rough probe on a rough sample, and SEM image of a RB_0.53 colloidal probe, scale bar = 500 nm. (c) RB_0.53 probe scanning at different locations on a 22 nm rough gradient substrate (top), friction loops (bottom) at 60 nN applied load for various $h/d$ roughness on the substrate ($h/d = 0.53$, black; $h/d = 0.39$, red; $h/d = 0.35$, green; $h/d = 0.29$, blue; $h/d = 0.23$, cyan; $h/d \approx 0$, magenta), scale bar = 200 nm. (d) Determination of $\mu$ from the measured friction forces and applied loads using the relation: $F_{friction} = \mu \cdot F_{load}$. (e) $\mu$ versus $h/d$ for a smooth probe (blue) and rough probes (red) on surfaces with various asperity size (12 nm ● (RB_0.45), 22 nm ■ (RB_0.53), 39 nm ♦ (RB_0.36) and 39&12 nm ▲ (RB_0.31 and smooth); ▼ (RB_0.25). (f) Correlation between $\mu$ and normalized packing packing fraction ($\phi^{RCP} = 0.64$, Inset: schematics of smooth particles sliding (right) and rough particles interlocking (left) under shear.

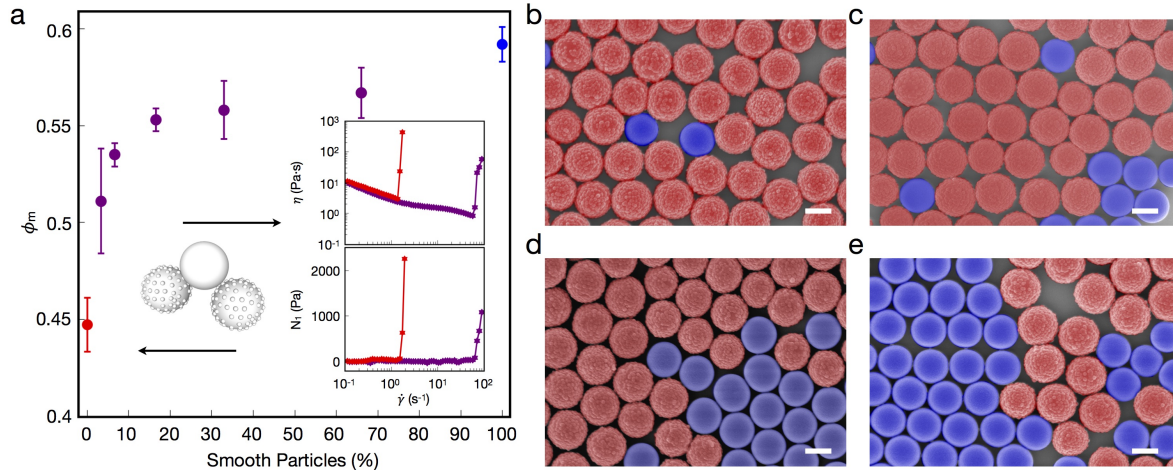

Figure 4. Engineering the rheological response using a tribological approach. (a) $\phi_m$ of RB_0.53 •, SM_0 •, and mixed SM_0 in RB_0.53 • as a function of mixing ratio. Inset left: schematics of a smooth particle breaking the interlocking between rough particles. Inset right: shear viscosity (top) and $N_1$ (bottom) versus shear rate for RB_0.53 ★ and 3.3 vol% of SM_0 in RB_0.53 ★ at $\phi$ = 44 %. SEM images of a suspension containing (b) 3.3 vol%, (c) 16.5 vol%, (d) 33.3 vol%, (e) 66.6 vol% of SM_0 (blue) in RB_0.53 (red). False colors, scale bars = 500 nm.

Methods

**Fabrication of Raspberry-like Silica Particles.** 69.4 mL of ethanol (analytical grade, ≥ 99.8 % Fluka, Switzerland), 11.4 mL of ammonia (25 % in water, Merck KGaA, Germany) and 9.2 mL of Milli-Q water (Merck Millipore, USA) were mixed by stirring at 400 rpm in a 200 mL Erlenmeyer flask. The flask had been pre-cleaned with KOH and HCl solutions to remove any possible nucleation sites from the flask. A 50 vol% mixture of TEOS (reagent grade 98 %, Fluka, Switzerland) and ethanol was injected into the mixed solution by a syringe pump (NE-1000, New Era Pump Systems Inc., USA) at a rate of 10 mL/hour. Different sizes of silica particles were produced by injecting different amounts of TEOS into the mixture solution. The injection time was between 45 to 60 min, aiming to synthesize particles with sizes between 550 to 700 nm (see Table SI). The purpose of producing a range of particle sizes was that different raspberry-like particles with different cores would have similar sizes after adsorption of the berries and silica heteronucleation. After the injection of the TEOS solution was finished, the mixture was kept stirring for 10 additional min. The suspensions were collected and centrifuged (Rotofix 32, Hettich, Germany) for 3 min at 4000 rpm. The supernatant solution was then removed, and the particles were resuspended in Milli-Q water. This procedure was repeated five more times to remove the initial reaction mixture and increase the monodispersity. After the final wash, the cleaned product was dispersed in Milli-Q water at concentration of 1 mg/mL.

The method to fabricate and tune particle surface roughness consists of two main steps: 1) the fabrication of raspberry-like particles via the electrostatic absorption of small silica particles on bigger silica cores and 2) a surface smoothing via a sol-gel route [4, 19].

In the first step, the synthesized smooth silica particles were used as the core particles. 400 mg of smooth silica particles and 250 $\mu$L of polydiallyldimethylammonium chloride solution (poly-DADMAC, 400-500 kDa, 20 wt% in water, Sigma-Aldrich, Switzerland) were dispersed in 240 mL of Milli-Q water. This aqueous suspension was stirred at 1000 rpm for 40 min. The positively charged poly-DADMAC molecules attached to the negatively charged (in neutral pH) silica particles surface via electrostatic attraction. The surface-modified particles were then washed three times (4000 rpm, 3 min), and the aqueous supernatant was exchanged with Milli-Q water to remove the excess of poly-DADMAC.

The surface-modified particles were positively charged and served as cores to electrostatically absorb unmodified smaller silica particles onto their surface. The $\zeta$-potential of smooth particles in 1 mM aqueous KCl solution was -61.6 ± 1.3 mV before modification and 17.8 ± 0.6 mV after being modified.

The cleaned surface-modified particles were dispersed (400 mg) in 120 mL of Milli-Q water. Smaller silica particles (1 wt%) of 39 nm (Klebosol, Clariant, France), 22 nm (Ludox® TM, DuPont, USA) or 12 nm (Corpusular, Microspheres-Nanospheres, USA) diameter were added to the suspension under stirring at 500 rpm for 60 min (details see Table S2). The obtained raspberry-like particles were washed three times by centrifugation (4000 rpm, 3 min), and the aqueous supernatant was exchanged with Milli-Q water. After the final wash, the cleaned raspberry-like particles were dispersed in 40 mL of Milli-Q water, giving a 1 wt% suspension.

In the second step, a sol-gel chemistry approach was applied to smoothen the surface of the raspberry-like particles. TEOS molecules were used, in order to grow a layer of amorphous silica on the surface of the raspberry-like particles through heterogeneous nucleation. 74.4 mL of ethanol, 12.2 mL of ammonia and 10 mL of 1 wt% raspberry-like particle suspension were mixed. A 5 vol% of TEOS in ethanol solution was injected by a syringe pump into the mixture with a rate of 2 mL/hour while sonicating. The TEOS solution was added in cycles. Each cycle consisted of adding 0.267 mL of TEOS solution following by an additional 25 min growing period without any injection. All smoothing experiments were performed under water cooling to keep the reaction temperature constant. The smoothed raspberry-like particles were washed three times (4000 rpm, 3 min), and the aqueous supernatant was exchanged with Milli-Q water. After the final wash, the cleaned particles were dispersed in 40 mL of Milli-Q water.

The smoothing layer has the primary effect of tuning the particles' surface roughness by modifying the height of the asperities, and secondly it strengthens the link between the core and the "berry-particles", preventing any possible detachment of the small particles from the cores, even at high shear. The extent of smoothing is controlled by varying the amount of the added TEOS solution. The desired theoretical thickness of the amorphous silica layer was calculated based on the estimated specific surface area of raspberry-like particles. The parameters of the rough particles produced in this work are summarized in Table SII.

**Characterization of Raspberry-like Silica Particles.** Scanning electron microscopy (SEM) samples were prepared by spreading particle suspensions on UV/ozone-cleaned (Pro-Cleaner PLUS, UVFAB, USA) silicon wafers. The samples were sputter-coated (CUU-010,

Safematic, Germany) with 4 nm of platinum to avoid surface charging prior to SEM imaging.

Atomic force microscopy (AFM, NanoWizard® NanoOptics, JPK Instruments AG, Germany) was used to record the surface topography and analyze the surface roughness. The AFM samples were prepared by spreading the particle suspensions on UV/ozone-cleaned round glass slides (AGL46R24-1, Agar Scientific, UK). The AFM images were recorded in AC mode by using BioLever mini (Olympus, spring constant = 26 N/m, Japan) and analyzed by a custom-written Mathematica (Wolfram Research, Inc., USA) code to derive their surface roughness. The AFM data were analyzed in four steps. First, each particle center was identified by recognizing the local maxima of a band-pass-filtered image. Second, the radius of each particle was extracted by fitting its profile with a sphere. Third, the spherical volume constructed using the fitted radius was subtracted from the AFM height image in order to obtain the true surface topography decoupled from the underlying curvature. In this way, we obtained a flattened particle surface. Finally, for the largest possible square inscribed in the scanned particle section, the centers of all asperities were identified by finding the local maxima, and the distances between one asperity and its neighboring asperities were recorded, so that the dimensionless roughness parameters $h/d$ could be calculated by knowing the average height of the asperities ($h$) and their average separation distance ($d$) as shown in Figure 1 (h).

**Centrifugation Experiments.** The centrifugation experiments were performed using both the smooth and the rough colloids. In order to prepare colloids with various initial volume fractions $\phi_i$, the following procedures was adopted. Initially, 1 mL of the 1 wt% suspension was centrifuged at 10000 rpm for 60 seconds and then 900 μL of the water was removed. Subsequently, another mL of suspension at 1 wt% was added to the sediment and the suspension was mixed, re-centrifuged and 1 mL of the supernatant was removed. Repeating this process, the sediment was progressively densified until the desired weight fraction was reached. The final 100 μL of suspension was kept in a 1.5 mL Eppendorf vial and sonicated for 30 min to redisperse the particles. The wt% was confirmed by measuring the dry mass using an ultra-microbalance (UMT2, Mettler Toledo, Switzerland). To convert the obtained wt% into vol%, a silica density of 1.8 g/cm$^3$ was used [34].

The prepared suspensions were sucked into rectangular capillary tubes (Length: 1 mm; Width: 0.05 mm; Height: 50 mm; Wall thickness: 0.05 mm, 5015, Vitrotubes$^{TM}$, USA) through capillary action. The filled capillaries were sealed with UV-curing glue (Norland

Optical Adhesive 63, Norland Products, USA) on glass slides and then centrifuged at 1500 rpm (acceleration ≈ 250 $g$) for 10 min. The sediment height, $H$, was recorded directly after centrifugation in order to calculate $\phi_m$. The sediment height divided by the tube's length, $L$, gives the sediment height in % (Figure 2 (a)). The sediment packing fraction is the slope of the linear fitting of the normalized sediment height ($H/L$) versus the suspension's initial volume fraction (Figure S1 (a)). After centrifugation, the samples were left under under the sole action of gravity. When no further evolution of the sediment height was observed (10 days), the sediment was assumed to be at its maximal compaction, corresponding to a volume fraction $\phi_f$ (Figure S1 (b)). This slow densification of the sediment is due to the combined effect of thermal fluctuations, which destabilize the force chains created by particle jamming under high shear in the centrifuge, and the action of gravity (the sedimentation length for our colloids is ≈ $100 \mu m \ll H$).

The reliability of the measured sediment volume fractions is further confirmed by the measured value of $\phi_f = 0.65$ for the SM_0 colloids, which is very close to the value of random-close-packing of monodisperse hard spheres $\phi^{RCP} = 0.64$. The small discrepancy may be due to the slight polydispersity of our SM_0 colloids.

**Shear-rheology Experiments.** The shear-rheology measurements were performed on a strain-controlled rheometer (ARES-G2, TA Instruments, USA) at 20 °C in a cone-and-plate geometry, using a 40 mm diameter stainless steel cone with an angle of 1°. The flow curves were recorded for increasing shear rates from 0.1 to 100 Hz. The rheological analysis was performed on the SM_0 suspensions from $\phi = 48$ % to 58 %, RB_0.53 suspensions from $\phi = 36$ % to 44 %, and $\phi = 48$ % for all other rough colloids.

The suspensions were prepared by centrifugation (4000 rpm, 15 min) to produce the sediments at $\phi_m$. Different amounts of Milli-Q water were added to the different sediments to adjust $\phi$ to the desired volume fraction. The suspensions were then sonicated for 30 min before performing the measurements.

**Fabrication of Model Rough Substrates.** 20 mm × 20 mm glass slides (Menzel-Gläser, Germany) were used as planar substrates to fabricate model rough surfaces. The glass slides were cleaned by ultrasonication for 10 min in toluene, 10 min in isopropanol, 10 min in ethanol and then 10 min in Milli-Q water, followed by 20 min UV/ozone cleaning (ProCleaner PLUS, UVFAB, USA). The cleaned glass slides were then immersed in 1 mg/mL polyethyleneimine (PEI, branched, high molecular weight, Sigma-Aldrich, USA) solution for 30 min while stirring at 200 rpm to switch the surface charge to positive. After the adsorption of PEI, the glass slides were rinsed with Milli-Q water and blown dry with a nitrogen jet.

In order to obtain a broad range of surface-roughness values on the same substrate, we created roughness gradients. The roughness-gradient samples were achieved by electrostatic absorption of single or binary populations of negatively charged silica particles onto the glass substrates. The gradients were prepared by dipping the PEI-coated substrates into the silica particle suspension using a linear-motion drive (Owis Staufen, Germany). 0.002 wt% of 12 nm suspensions, 0.004 wt% of 22 nm suspensions, and 0.004 wt% of 39 nm suspensions in Milli-Q water were used. The immersion profile was set to $z(t) = 6.94 \cdot 10^{-6} \cdot t^2$, where $z$ [mm] is the position on the gradient at time $t$ [s]. The maximum immersion time was 1200 s. The settings were such that during the 20 min immersion a gradient with 10 mm length was prepared. At the end of the process, the samples were immediately removed from the suspension and rinsed with Milli-Q water [35]. The binary gradients were dipped in a 12 nm-particle suspension for 10 min after coating with the 39 nm particles.

An amorphous silica layer was grown on the surface, mimicking the smoothing layer on the raspberry-like particles' surface. In detail, 7.44 mL of ethanol, 1.22 mL of ammonia and 1 mL of Milli-Q water were mixed in a polytetrafluoroethylene (PTFE) container, and 600 $\mu$L of 1 vol% of alcoholic TEOS solution mixture was added. A particle-coated substrate was placed in the mixture for 30 min while stirring at 200 rpm. After the reaction, the substrate was rinsed with Milli-Q water and dried with a nitrogen jet. The fabrication process is shown in Figure S2.

**Friction Measurements on Roughness-gradient Samples.** The friction on planar samples was measured by AFM in contact mode via LFM. In the LFM mode, the torsion of the AFM cantilever is measured by a photodetector. During lateral scanning, the cantilever experiences a torque due to friction. The lateral deflection of the cantilever is directly proportional to the frictional forces. The AFM records the lateral deflection in volts by computing the difference between the quadrant photodetector signals. This value can be converted to force units by an appropriate calibrations of lateral sensitivity of the photodetector and the torsional spring constant of the cantilever. The measured lateral deflection versus scanning distance plots is referred to as a friction loop, and consists of trace and retrace curves indicating the forward and backward motion of the cantilever during the scanning (see Figure 3 (c) and Figure S4, S5, S6, S7).

The measurements were carried out using both smooth silica colloidal probes (sphere radius = 300 nm, Novascan Technologies, USA) and rough colloidal probes on rough gradient replicas. The rough colloidal probes were fabricated by attaching a raspberry-like silica particle to the end of a tip-less AFM cantilever (NSC36/Tipless, MikroMasch, Estonia) with a home-built micro-manipulator coupled to a microscope (BX 41, Olympus microscope, Japan). In a first step, a small amount of Araldite epoxy glue was picked up with a sharpened tungsten wire (Wire.Co.UK, UK) and smeared on the end of the cantilever using the micro-manipulator. In a second step, the required raspberry-like particle was picked by another sharpened tungsten wire and was placed precisely over the glue. The smooth colloidal probe and the rough colloidal probes are shown in Figure S3.

The friction measurements were performed in a custom-made liquid cell using Milli-Q water as the medium to mimic the conditions of the colloids in the rheological experiments accurately. The samples were fixed in a 40 mm Petri dish (93040, TPP, Switzerland) using UV-curing glue (Norland Optical Adhesive 81, Norland Products Inc., USA) and immersed in Milli-Q water before the measurements. The samples and the probes were UV/ozone cleaned for 15 min prior to the friction measurements.

The vertical sensitivity (nm/V) of each probe in air was determined by performing a force curve on a hard silica surface. Then the normal spring constant (N/m) of the cantilever was determined by the thermal-noise method [36]. The sensitivity was corrected to the refractive index change of the liquid before staring the measurement.

The calibration of the lateral sensitivity and torsional spring constant of the cantilever was performed after the friction measurements. The lateral calibration constant, i.e., the constant to convert the lateral voltage signal of the AFM to friction force, was determined scanning a wedge-shaped calibration grating (TGF11 grating by MikroMasch, Tallinn, Estonia) immersed in Milli-Q water. The grating has one-dimensional arrays of trapezoidal steps and the sidewalls and the horizontal top surfaces form an angle of 54.74°. Since our colloidal probes are too small to be used with the standard grating, a 4.28 $\mu$m silica particle (SiO2- R-4.0, Microparticles GmbH, Germany) was glued on the same cantilever used for the measurement using a micro-manipulator in order to perform the calibration (See Fiugre S8). Trace and retrace line scans (friction loops) of 128 pixels, corresponding to 3 $\mu$m were collected at 1 Hz on the grating at increasing loads between 20 nN and 120 nN. From these scans, the cantilevers were calibrated following the procedure of [37, 38], after measuring the particle radius and the cantilever thickness for each probe using an SEM.

The friction-coefficient measurements over the model rough samples were carried out by scanning the substrate in the fast-scan direction orthogonal to the cantilever axis. The scan area and scan rate were fixed at 2 $\mu$m $\times$ 100 nm (1024 px $\times$ 51 px) and 1 Hz, respectively. The friction measurements were carried out from the roughest region toward the smooth end on the roughness-gradient samples. At every scanning area, the friction loops were recorded at different applied loads $F_{load}$ from 10 nN to 60 nN. The average friction force $F_{friction}$ for a given load was obtained by averaging half the vertical difference between the trace and retrace curves for all the friction loops. The friction coefficient $\mu$ is extracted as $F_{friction} = \mu \cdot F_{load}$, as shown in Figure 3d.

**Ball-dropping Tests.** In order to produce a visual demonstration of the different shear-thickening response of the smooth and rough particle suspensions, we performed ball-dropping tests.

SM_0 suspensions of 48%, 54%, 57%, and 58% (Videos S1, S2, S3, and S4) as well as RB_0.53 suspensions of 36%, 40%, 42% and 44% (Videos S5, S6, S7, and S8) were prepared and poured into polystyrene cuvettes (BRAND, Germany). A high-speed camera (FASTCAM SA-Z, Photron, USA) was used to record image sequences (frame rate = 5000 fps) of a steel ball (diameter = 2 mm, weight = 45 mg) dropping from 20 cm height onto the surface of the suspension. The estimated shear rate at impact was $\approx$ 400 Hz, i.e. in the ST regime for all suspensions.

**Author Contributions** L.I. and N.D.S conceived the study. C-P.H. performed all experiments. M.Z. contributed to the fabrication of the rough colloids. S.N.R. contributed to the LFM experiments. All authors discussed and analyzed the data and wrote the manuscript.

ACKNOWLEDGMENTS

C-P.H. and L.I. acknowledge financial support from the Swiss National Science Foundation grant PP00P2 144646/1 and the ETH Zurich research grants ETH-49-16-1. We thank Jan Vermant for fruitful discussions. We thank Svetoslav Anachkov for providing the Mathematica code, Thomas Schweizer for assistance with shear rheology measurement, Rebecca Huber for assistance with gradient sample preparation, Christopher McLaren for help with high-speed video recording and Andre Studart for the SEM access.

# Roughness-Dependent Tribology Effects on Discontinuous Shear Thickening: Supplementary Information

Chiao-Peng Hsu,[1,2] Shivaprakash N. Ramakrishna,[2] Michele Zanini,[1] Nicholas D. Spencer,[2] and Lucio Isa[1, *]

[1]*Laboratory for Interfaces, Soft Matter and Assembly,*
*Department of Materials, ETH Zurich, Zurich, Switzerland.*
[2]*Laboratory for Surface Science and Technology,*
*Department of Materials, ETH Zurich, Zurich, Switzerland.*

---

* lucio.isa@mat.ethz.ch

Table SI. Injection Conditions for the Synthesis of Core Particles

| Injection Time (min) | 45 | 50 | 55 | 60 |
|---|---|---|---|---|
| Particle size (nm) | 574±21 | 627±18 | 658±29 | 678±24 |

Table SII. Details of the Smooth and Raspberry-like Silica Particles

| | SM_0 | RB_0.25 | RB_0.31 | RB_0.36 | RB_0.45 | RB_0.53 |
|---|---|---|---|---|---|---|
| Core size (nm) | 678±24 | 658±29 | 574±21 | 627±18 | 547±21 | 658±24 |
| Berry size (nm) | - | 39±2 & 12±1 | 39±2 & 12±1 | 39±2 | 12±1 | 22±2 |
| Amount of berries added (mL at 1 wt%) | - | 8.6 & 0.25 | 10 & 0.28 | 9 | 0.64 | 6.7 |
| Smoothening layer (nm) | - | 15±2 | 10±1 | 10±1 | 10±1 | 10±1 |
| h/d | ≈ 0 | 0.25±0.03 | 0.31±0.04 | 0.36±0.04 | 0.43±0.05 | 0.53±0.05 |

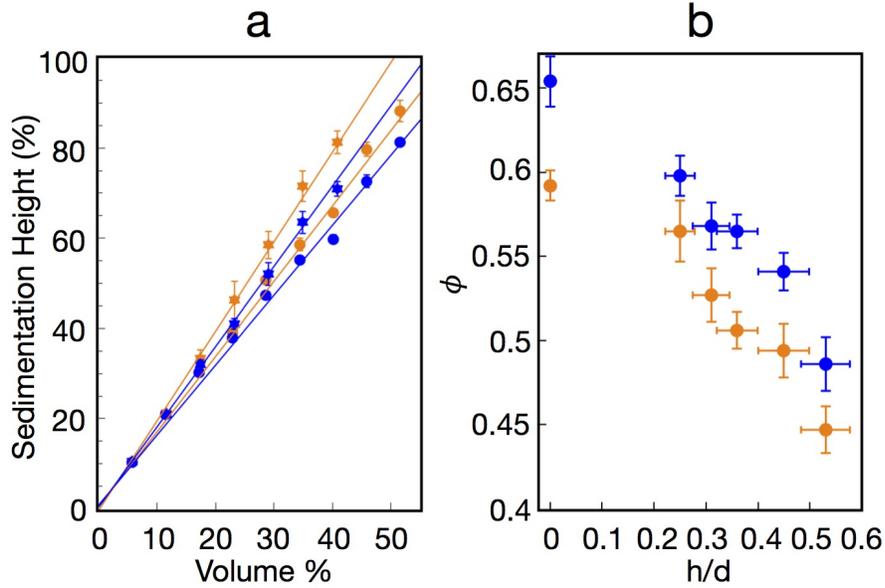

Figure S1. (a) Determination of $\phi_m$ (orange) and $\phi_f$ (blue) of rough colloids ★ (RB_0.53) and smooth colloids ● (SM_0). (b) $\phi_m$ (orange) and $\phi_f$ (blue) of the colloidal suspensions with different surface roughness expressed in terms of $h/d$.

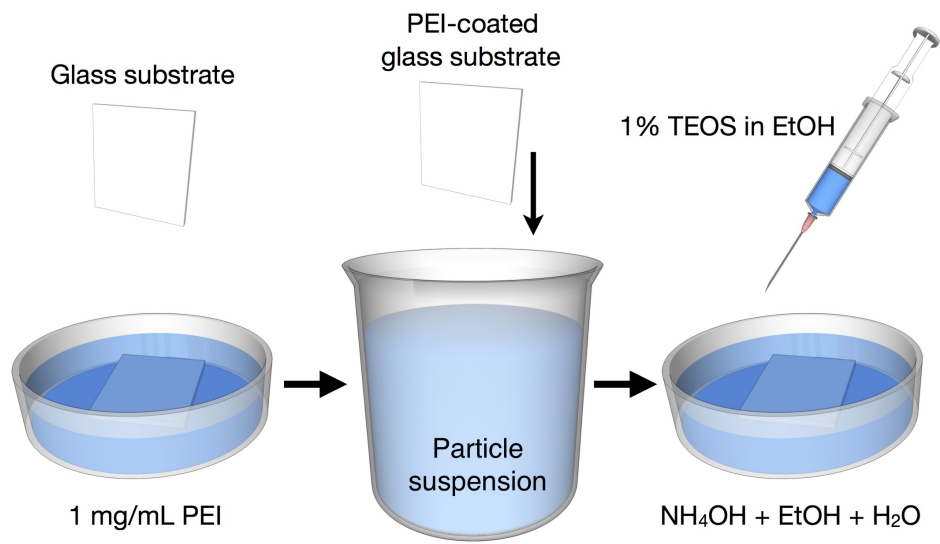

Figure S2. Schematics of the fabrication of roughness gradient samples.

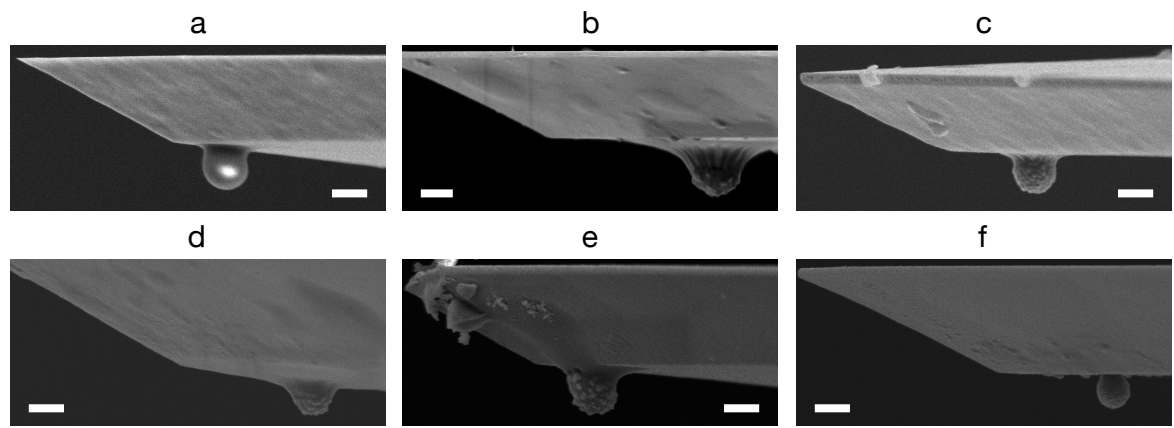

Figure S3. SEM images of a (a) smooth colloidal probe, (b) RB_0.25 probe, (c) RB_0.31 probe, (d) RB_0.36 probe, (e) RB_0.43 probe, (f) RB_0.53 probe, scale bars = 500 nm.

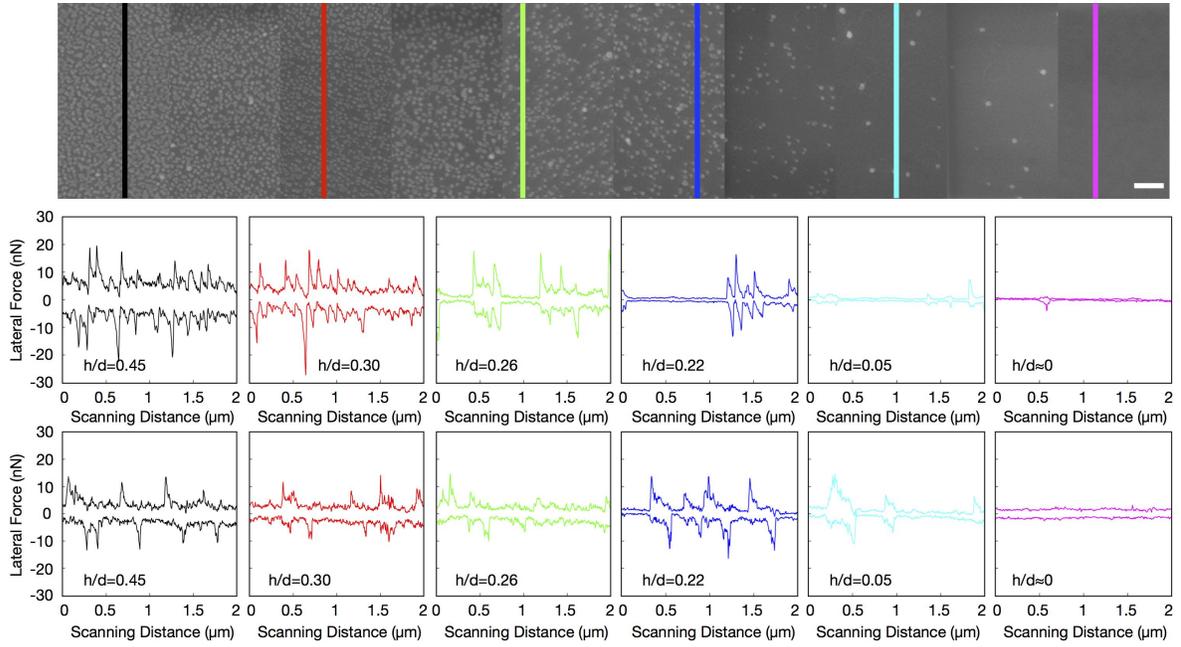

Figure S4. Friction loops obtained between a RB_0.45 probe (top) and a smooth colloidal probe (bottom) scanning at different locations on a 12 nm roughness gradient substrate at 60 nN applied load for various $h/d$ roughness on the substrate ($h/d = 0.45$, black; $h/d = 0.30$, red; $h/d = 0.36$, green; $h/d = 0.22$, blue; $h/d = 0.05$, cyan; $h/d \approx 0$, magenta), scale bar = 200 nm.

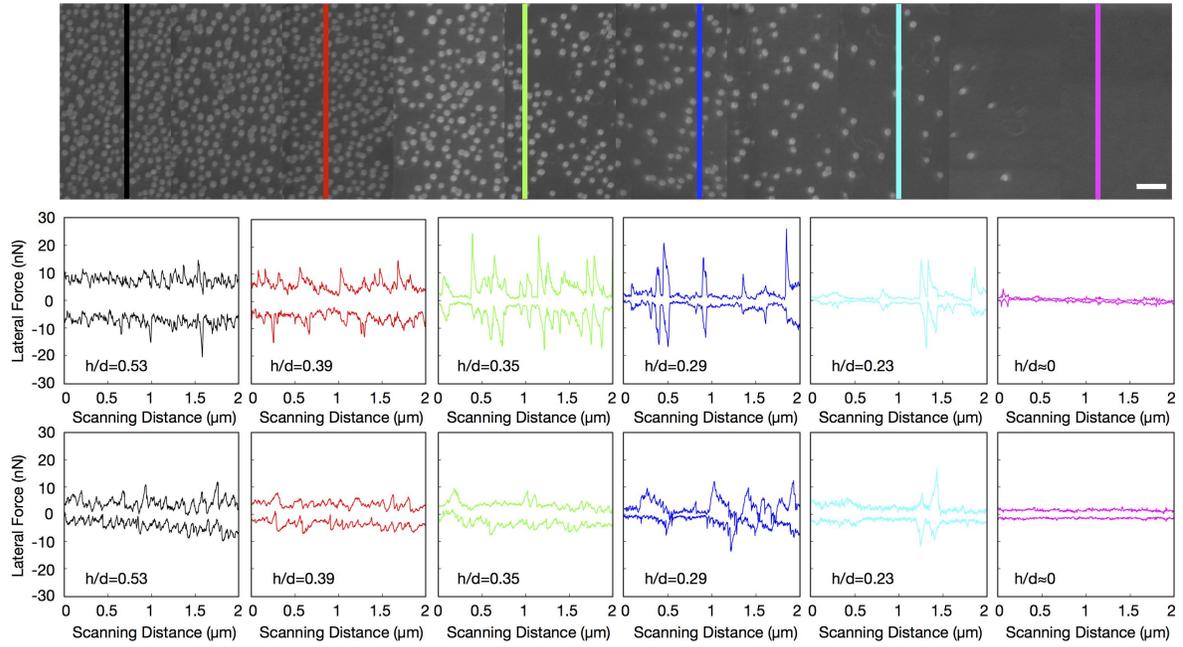

Figure S5. Friction loops obtained between a RB_0.53 probe (top) and a smooth colloidal probe (bottom) scanning at different locations on a 22 nm roughness gradient substrate at 60 nN applied load for various $h/d$ roughness on the substrate ($h/d = 0.53$, black; $h/d = 0.39$, red; $h/d = 0.35$, green; $h/d = 0.29$, blue; $h/d = 0.23$, cyan; $h/d \approx 0$, magenta), scale bar = 200 nm.

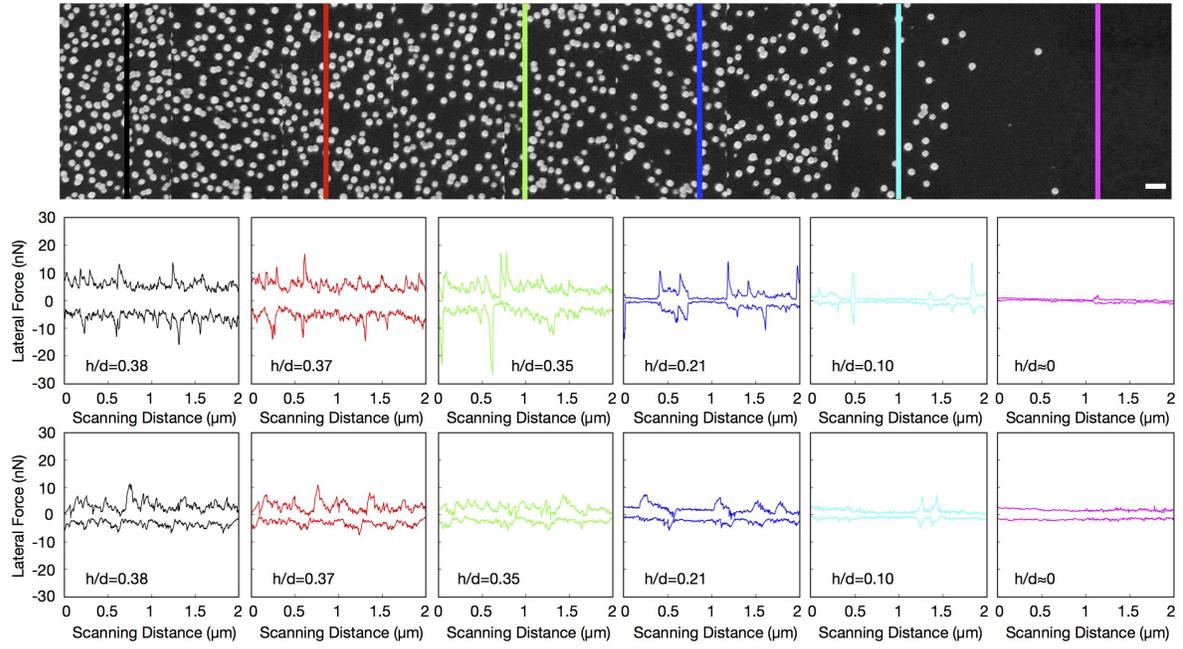

Figure S6. Friction loops obtained between a RB_0.36 probe (top) and a smooth colloidal probe (bottom) scanning at different locations on a 22 nm roughness gradient substrate at 60 nN applied load for various $h/d$ roughness on the substrate ($h/d = 0.38$, black; $h/d = 0.37$, red; $h/d = 0.35$, green; $h/d = 0.21$, blue; $h/d = 0.10$, cyan; $h/d \approx 0$, magenta), scale bar = 200 nm.

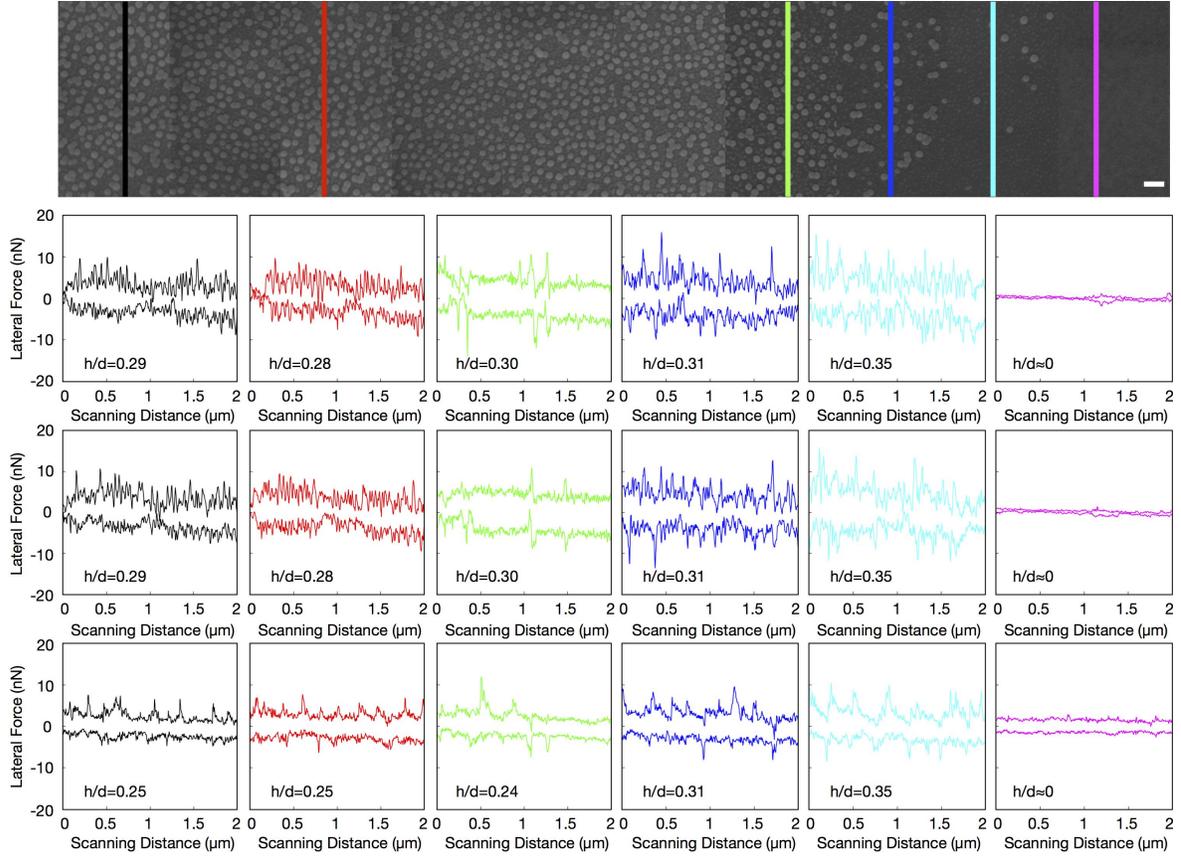

Figure S7. Friction loops of a RB_0.25 probe (top), a RB_0.31 probe (middle) and a smooth colloidal probe (bottom) scanning at different locations on a 39&12 nm rough gradient substrate at 60 nN applied load for various $h/d$ roughness on the substrate ($h/d = 0.29$ (RB_0.25 and RB_0.31 probe), 0.25 (smooth probe) black; $h/d = 0.28$ (RB_0.25 and RB_0.31 probe), 0.25 (smooth probe), red; $h/d = 0.30$ (RB_0.25 and RB_0.31 probe), 0.24 (smooth probe), green; $h/d = 0.21$, blue; $h/d = 0.10$, cyan; $h/d \approx 0$, magenta), scale bar = 200 nm. The difference of $h/d$ at the first 3 locations is due to the fact that a 600 nm smooth probe cannot experience the small asperities when the big ones are close enough. This results in a smaller effective $h/d$ when using a smooth probe to scan these locations.

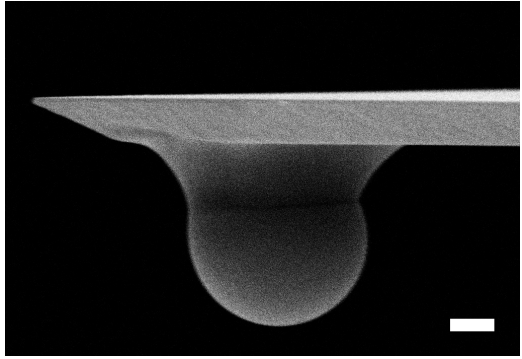

Figure S8. SEM image of a 4.28 µm silica particle glued over a RB_53 colloid on the same cantilever after the latter was used for friction measurements in order to carry out the lateral force calibration, scale bar = 1 µm.

**Video S1**: Video of a steel ball dropping on 48% SM_0 suspension

**Video S2**: Video of a steel ball dropping on 54% SM_0 suspension

**Video S3**: Video of a steel ball dropping on 57% SM_0 suspension

**Video S4**: Video of a steel ball dropping on 58% SM_0 suspension

**Video S5**: Video of a steel ball dropping on 36% RB_53 suspension

**Video S6**: Video of a steel ball dropping on 40% RB_53 suspension

**Video S7**: Video of a steel ball dropping on 42% RB_53 suspension

**Video S8**: Video of a steel ball dropping on 44% RB_53 suspension